# A Resource Intensive Traffic-Aware Scheme for Cluster-based Energy Conservation in Wireless Devices

Marios C. Charalambous
Computer Science Department,
University of Nicosia
46 Makedonitissas Avenue, 1700
Nicosia, Cyprus
chalambous.mar@st.unic.ac.cy

Constandinos X. Mavromoustakis
Computer Science Department,
University of Nicosia
46 Makedonitissas Avenue, 1700
Nicosia, Cyprus
mavromoustakis.c@unic.ac.cy

Muneer Bani Yassein
Jordan University of Science and
Technology,
P.O. Box 3030,
Irbid 22110, Jordan
masadeh@just.edu.jo

*Abstract*—Wireless traffic that is destined for a certain device in a network, can be exploited in order to minimize the availability and delay trade-offs, and mitigate the Energy consumption. The Energy Conservation (EC) mechanism can be node-centric by considering the traversed nodal traffic in order to prolong the network lifetime. This work describes a quantitative traffic-based approach where a clustered Sleep-Proxy mechanism takes place in order to enable each node to sleep according to the time duration of the active traffic that each node expects and experiences. Sleep-proxies within the clusters are created according to pairwise active-time comparison, where each node expects during the active periods, a requested traffic. For resource availability and recovery purposes, the caching mechanism takes place in case where the node for which the traffic is destined is not available. The proposed scheme uses Role-based nodes which are assigned to manipulate the traffic in a cluster, through the time-oriented backward difference traffic evaluation scheme. Simulation study is carried out for the proposed backward estimation scheme and the effectiveness of the end-to-end EC mechanism taking into account a number of metrics and measures for the effects while incrementing the sleep time duration under the proposed framework. Comparative simulation results show that the proposed scheme could be applied to infrastructure-less systems, providing energy-efficient resource exchange with significant minimization in the power consumption of each device.

## I. INTRODUCTION

The consideration of mechanisms that enable data availability along with the energy consumption in mobile devices in today's wireless deployments is becoming more timely, as wireless resource-sharing applications are rapidly gaining ground. A variety of device-dependent applications were born, figuring the necessity for developing a scheme for conserving energy, whereas, it allows resources availability. As the energy efficiency mechanism is of primary importance for the end-to-end communication in wireless peer-communicating devices, the underlying supporting mechanisms should ensure that nodes transmit their information with minimum energy consumption. In resource-sharing applications the delay durations cannot be guaranteed, whereas the nodes that are hosting the resource sharing processes are consuming more energy [15] due to high activity nature of the wireless nodes. Notwithstanding the EC should be severely balanced in such applications, there is a need to apply, to the existing resource-sharing in peer-applications, a traffic-aware mechanism. This mechanism will encompass the manipulation of resource sharing process in a reflective way and enabling a direct impact on EC of each device. This reflective scheme has to take into consideration the continuity of the incoming traffic and the communication (data and control packets) among peers in order to provide the energy conservation schedules of the wireless devices.

Power control in wireless devices aims to determine the transmit periods and the associated power level such that the energy consumed is steadily reduced, whereas at the same time it aims to guarantee the resource sharing stability. Since nodes in wireless networks typically rely on their battery energy, this work, proposes a mechanism which hosts a traffic-aware scheme for conserving energy in wireless environments using a Clustered-based mechanism. This mechanism is encapsulated in a middleware and encompasses a scheme which evaluates the scheduled activity periods of each node, in order to measure the scheduled time that each node can safely sleep. The proposed framework uses a quantitative model based on the supervision of the Sleep-Proxy (a role-based node) which actively, in a clustered-environment, takes into consideration the cooperative caching and the resource exchange scheme for enabling Energy Conservation in the Cluster.

In this work, each node uses different assignment of sleep-wake schedule estimation, based on the minimum pair-wise EC and the derived backward difference of the estimated traffic introduced in III. This is performed in order to enhance node hibernation and avoid mutation which will result in network partitioning. The Sleep-Proxy nodes are utilized in the cluster, for enabling mitigation between the energy consumption and the overall activity periods of the nodes. Simulation experiments and study were carried out for the energy conservation and the evaluation of the proposed model taking into account a number of metrics hosted by the proposed scheme and for estimation of the effects of incrementing the sleep time duration to conserve energy.

The organization of the paper is as follows: Section 2 discusses the related work that has been done on EC featuring out the basic energy conservation principles and conducted solutions by similar schemes. Section 3 then introduces the proposed middleware which hosts the mechanisms for the Clustered Sleep-Proxy nodes and Scheduled Activity Periods apparatus for delay sensitive streaming considering the sleep time duration assignment of each node, followed by Section 4 which provides the evaluation and comparative simulation results of the proposed scheme in contrast to the energy consumption associated with the proposed middleware. Finally, Section 5 concludes with a summary of our contribution and suggestions further research.

## II. RELATED WORK

By prolonging the lifetime of wireless devices, a broader field of applications can benefit from using a Peer-to-Peer technology in a mobility-based framework. Content delivery, media streaming [1], games [2], and collaboration tools are examples of applications fields that use Peer-to-Peer networks today. These applications can only be implemented with guaranteed QoS support, using wired environments' support. This is primarily the reason that the number of applications beyond file sharing is kept on a low implementation level, despite the general character of Peer-to-Peer networking. The type of application in wireless devices typically rely their presence on the energy that the devices host. Therefore, a supporting middleware that employs the energy consumption problem with contrast to the incoming traffic on each node is considered very important. Recent research has already addressed the MP2P connectivity from different perspectives. Proem [3] is a mobile middleware providing solution for developing and deploying applications for mobile ad hoc networks. In Proem, middleware is responsible for presence and discovery services as well as being an identity, data space and community manager. Proem has been designed for mobile peers in ad hoc networks whereas in Mobile Chedar [4] also peers with fixed P2P network connections are supported. Traditional middleware schemes applied in mobile computing such as the ALICE [6], have been used to provide support for client-server architectures in nomadic environments. Moreover, RPC-based middleware has been enhanced with queuing delaying or buffering capabilities in order to cope with intermittent connections. Examples of these behaviors are Rover [7] and Mobile DCE [8-10].

In addition to the existing architectures, a fertile ground of the development of new approaches has been the association of different parameters with communication mechanisms in order to reduce the energy consumption. These mechanisms can be classified into two categories: Active and passive schemes. Active techniques conserve energy by performing energy conscious operations, such as transmission scheduling using a directional antenna [11], and energy-aware routing [12-13]. Passive techniques conserve energy by scheduling the interfaces of the devices to the sleep mode when a node is not currently taking part in communication activity [14] and host different adaptive methodologies like the Adaptive Dynamic Caching Energy Conservation (ADCEC) [15] which takes into consideration the traffic pattern that a node is experiencing as incoming and outgoing traffic. Authors in [15] consider the association of EC problem with different parameterized aspects of the traffic (like traffic prioritization) and enable a mechanism that tunes the interfaces' scheduler to sprawl in the sleep state according to the activity of the traffic of a certain node in the end-to-end path. Additional mechanisms like the LEACH proposed in [16], aim to minimize energy consumption in WSNs through a cluster-based operation. In this work the proposed framework deals with the traffic awareness in a self-adaptive way for each node while the end-to-end communication takes place. The main goal of the proposed scheme is to minimize the energy consumption for the communication among nodes which are exchanging resources. This is achieved by estimating the residual energy of the nodes, where nodes with high residual energy are selected as CHs. This work on the contrary with the LEACH mechanism used in [16] and taking into account clustering considerations in [17], assigns the role of Sleep-Proxy (SP) to nodes according to a maximum Energy stored for time duration $t$ in the cluster, in a decentralized estimated manner. The proposed scheme provides scalability by using a Cluster-Head proxy which limits the number of transmissions between nodes, thereby enabling a higher number of nodes to be actively deployed in the network. The Sleep Proxy which is assigned, provides local activity information using intra-cluster information exchange, and assigns the next sleep-time duration to each node according to the aggregated traffic that it was destined for each node in a period of time. The next section presents the proposed resource intensive traffic-aware scheme and the middleware hosting the associated mechanisms. These mechanisms are aiming the estimation of the next sleep-time duration for each node by enabling a traffic-aware sleep schedule assignment. Through the traffic-aware scheme the energy consumption can be significantly reduced and network lifetime can be further prolonged.

## III. TRAFFIC-AWARE MIDDLEWARE SCHEME FOR CLUSTER-BASED ENERGY CONSERVATION

### A. Traffic-driven Middleware and supported mechanisms

Traffic-aware policy requires an active scheme to be applied, through which, the traffic will reflect a certain impact on the nodes taking into account the EC trade-offs. Wireless devices should consider the incoming traffic, in order to adapt and reflect a certain feedback to the EC mechanism. A Middleware which hosts traffic changes and has a direct impact through the estimated scheme presented in Section III.B, is shown in Figure 1. Figure 1 shows a cross layer interaction through a mechanism for traffic-awareness in an end-to-end manner. In particular, real-time media traffic such as voice and video typically have high data rate requirements and stringent delay constraints, whereas wireless nodes generally have limited or momentarily connectivity. The proposed middleware enables the data packets to be traversed and manipulated through the utilized Data Link, Network, and Transport layers by considering the traffic awareness mechanism and the model for volume estimation to be reflected on these layers. The proposed traffic-aware mechanism evaluates (after the bootstrap process of the system) the estimated (quantified as Volume/Capacity) traffic that is destined for each node. In this way it enables –through the proposed mechanism- estimation for the next slot sleep duration of the node. This evaluation is performed in an interactive way through the mechanism in Section III.B. These mechanisms are performed in order to tune the wireless interface of each device to sleep/wake according to the activity of each individual device in the resource exchanging path. Packet classification methodology was utilized as in [15] in order to mark the packets that are exchanged whether they are delay sensitive or not. In turn, if packets are considered as delay sensitive, strict deadlines are applied by the sender, according to the specifications set in the network. In the case where packet deadlines cannot be satisfied, then cached packets of nearby nodes, enable recovery using the promiscuous caching [18]. This mechanism enables the resources' replication and increases the resource sharing reliability [18]. The quantitative mechanisms shown in Figure 1, are depicted in the following sections with the quantitative analysis.

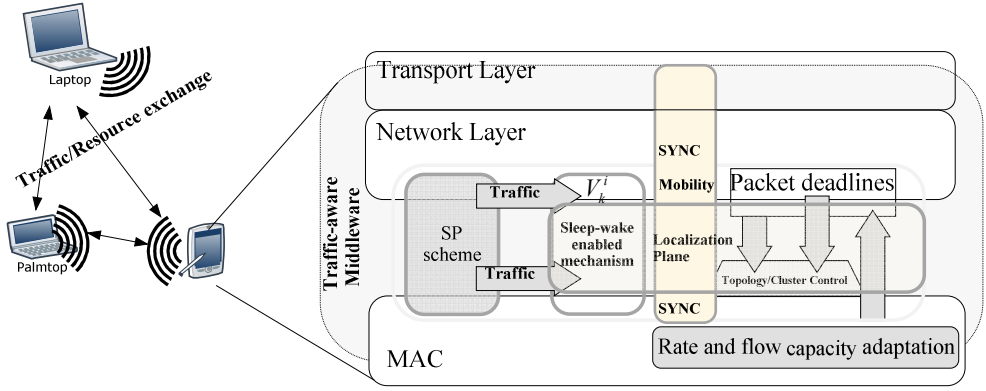

Figure 1. The traffic-aware interactive middleware mechanism with the associated influenced layers in the communication stack.

## B. Clustered Sleep-Proxy nodes and Scheduled Activity Periods mechanism

With event-based applications, network traffic is sporadic and connectivity (activity) is not required at all times. In other words, the nodes are in sleep state until an incoming traffic will enable them to get into the active state, whereas, if resources (sharing pieces) were missed they can be stored onto intermediate nodes and forwarded to destination, when destination node becomes active and available. Predetermined or proactive activity scheduling may decrease the network lifetime as it enables active-time uniformity, where, nodes are in active state for period of time that it is not necessary and beneficial. The activity period(s) of a node is primarily dependent on the nature and the spikes of the incoming traffic destined for this node [19]. If transmissions are performed on a periodic basis then the nodes' lifetime can be predicted/estimated. The network lifetime in the presence of periodic sleep-wake periods, even in the presence of traffic or not, will be significantly decreased [19] due to frequent and -unneeded- activity. Assuming that a number $n$ of nodes are set in a landscape Q whose edges correspond to the reliable communication links between any pair of $n$ nodes, if exist. This work formally defines a graph $G = (N, e)$ with a set of nodes $N=\{1, 2, \ldots, n\}$ and a set of edges $e$, where $|N| = n < \infty$. If a node $i \in N$ can reliably communicate with other node $j \in N$, then an edge between wireless nodes i, j exists, i.e., $(i, j) \in e$ $(i \neq j)$. G is an undirected and connected graph, the communication link between nodes $i, j$ is symmetric and there is at least one (routing) path from each node to every other node(s). Let $N(i)=\{j \in N : (i, j) \in e\}$ be the set of neighbors of node $i \in N$ and $d(i) =|N(i)|$, i.e., $d(i)$ is the degree of node i. Let $CH_n$ be the Cluster Head for a set of nodes $N =\{1, 2, \ldots, n\}$ selected according to the pair-wise algorithm of pseudocode in Figure 2. In Figure 2 the $E_{max}(i, j)$ is the maximum energy of either $i$ or $j$ before coming in contact, and $E_{remaining}$ is the remaining energy of the pair $i+j$ at time T, estimated with $E_{remaining} = E_i + E_j$. This procedure takes place during the nodes intermeeting duration denoted as $\tau$. In a certain cluster the $n-1$ nodes remaining, are re-evaluated in order to be ranked for time $\tau$ for potentially becoming a CH, according to the maximum-residual energy of each node. CHs are the organization leaders of a cluster and are required to organize communication activities in the cluster. The CHs are responsible to assign the sleep-Proxy (SP) to a certain node in the cluster and oversee the SP. This assignment occurs according to the Equation 1 where the SP activity is undertaken by the node with the highest probability as follows:

$$SP_L = C_L \cdot \frac{E_{residual}}{E_{max}}, \forall n \in N \quad (1)$$

where $C_L$ is the probability of the node to become a CH (Figure's 2 selection algorithm) and $SP_L$ is the probability of the node to act as a SP. Every node can become a SP only when the cluster has no other available node to act as a SP. Each selected node by the CH acts as a SP for one round T. The SP selection algorithm is executed during the set-up phase of each round, and guarantees that each node will become a SP at some time during the network's lifetime. The CH rotates the role of SP on a round-by-round basis in a fair manner and according to Equation 1, in order to enable nodes' energy to be consumed evenly.

```
do
   for 1 to n in the path P
      find the pair in P for which
         E_remaining > E_max(i, j)
         // E_remaining = E_i + E_j
   if  E_i > E_j
         i becomes CH
   else
         j becomes CH

   for 1 to n-1 in path P
      find  N_CH = Max(E_{n-1}) ∀n ∈ N
while ( T < τ )
set_CH ( N_CH )
```

Figure 2. Selection algorithm of CH according to the remaining Energy of pair-wise communicating nodes.

The steps that are followed by the CH and the Sleep-Proxy role (that is assigned to nodes in the cluster) are:
1. Selection of the CH in the formed cluster (pairwise Energy comparison algorithm of Figure 2).
2. CH assigns the SP role to a node, according to the Equation 1.
3. SP then enables the mechanism of the traffic backward-difference estimation (Section C).
4. The schedule of the transmission is being handled through the assigned active periods of the node that the data is to be destined.

## C. Intercluster Sleep Proxy scheme based on incoming nodal traffic for EC

SP assigns the schedules for each node according to the remaining energy and the duration of the activity periods of each node. This enables the node to transmit or receive (TX/RX mode) when in active state and being with inactive TX/RX interfaces when set in sleep state. The SP can select a node and assign the node to get into the idle and sleep state respectively through the estimations of the activity periods of the previous moments (backward estimation) as shown in equations 2.1-2.2, 3.1 and 4. If node's location has changed or the sleep period and/or $t_{sleep}$ expires, then node is inserted into the idle and active state respectively. The 'listen' and 'sleep' intervals of the proposed scheme are shown in figure 3. It is important to mention that the SP will guide the intra-cluster sleep-wake schedule with reference to each nodes' sleep assignments, and in turn nodes' sleep assignments will increase or decrease the activity moments of each node.

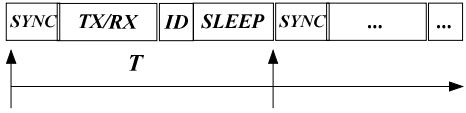

Figure 3. Listen and sleep intervals of the proposed scheme.

### 1) Pairwise active-time comparison and characterization

Idle time is evaluated by comparing the idle time durations of nodes using a pairwise notation. Let $t_{N_A}$ be the active time for node A in the interval [0..T] and $t_{N_B}$ be the active time for node A in the interval [0..T]. In the pairwise comparison two nodes A, B which have direct contact can be compared using current and previous moments. The next time duration is a function of the current time duration being active $S_{N_{A(\tau)}}$ and the previous time slot in $[0..T]^k$, for $k$ slots in the interval $0..T$, for nodes $A$ and $B$, it stands that:

$$\nabla t_{N_A} = \nabla S_{N_A} = S_{N_{A(\tau)}} - S_{N_{A(\tau-1)}} \quad (2.1) \text{ and}$$

$$\nabla t_{N_B} = \nabla S_{N_B} = S_{N_{B(\tau)}} - S_{N_{B(\tau-1)}} \quad (2.2)$$

If $t_{N_A} > t_{N_B}$, then if there is no incoming traffic destined for node A, it subsequently gets in the idle state and informs the SP of the cluster. Equations 2.1 and 2.2 show the time-oriented backward difference traffic evaluation scheme, where, the difference of the traffic that was destined for a node can affect the sleep-time of a node –measured model in Equation 4-enables EC. Hence, in order to evaluate the idle time $t_{idle}$ for each node that will get into the idle state, the following estimation takes place:

$$t_{idle} = \frac{(T - \max(d_p))}{n} \quad (3.1)$$

where $d_p$ is the maximum delay in the end-to-end path from a source to a destination where the reference (i.e. A) node lays in, $T$ is the round/cycle for which $t_{idle}$ is evaluated, and $n$ is the number of hops. $d_p$ is calculated as:

$$d_p = \sum_{i=0}^{i-1} \delta_i + T_i \quad (3.2)$$

where $\delta_i$ is the duration where the requested data was hosted onto $i$-node, and $T$ is the transmission delay.

### 2) Inter-cluster Node's sleep-time

Within the formed cluster, the sleep-time is evaluated by taking into account the delay of the stored data-destined for the node which is set in the sleep state-onto any other node. If a node is in the sleep-state and this node has data packets to receive, then these packets in order to be available and recoverable can be stored temporarily onto nearby hop-neighboring (active) nodes. Based on this scenario let $t_\delta^i$ be the delay that the stored packets are experiencing onto i-node (nearby neighboring node to destination) and $V_k^i$ the volume of traffic destined for node k and stored onto node $i$, then the $T_{sleep}(t)$ is measured by:

$$T_{sleep}(t) = \left( \frac{\sum_{i=0}^{k} C_{i \to k}(t) - \sum_{i=0}^{k} V_k^i}{\sup \sum_{i=0}^{k} C_{i \to k}(t)} \right)^n \cdot d_p \quad (4)$$

where $T_{sleep}(t) < T$, $T_{sleep}(t) < t_\delta^i$, where $C_{i \to k}(t)$ is the allowed capacity of the channel in the path p, $d_p$ is the maximum delay in the end-to-end path from a source to a destination. If the outsourced data volume $V_k^i$ is high, then $T_{sleep}(t)$ becomes smaller for active incoming traffic if node is still in the sleep state. This mechanism enforces the node to sleep if there are no incoming packets destined for the node being in the sleep state. If a node experiences a long $T_{sleep}(t)$ then this node gets into the active state since it stands that $T_{sleep}(t) < T$, $T_{sleep}(t) < \delta_p$.

The SP's sleep time, as a role-based node is evaluated in the [0..T] according to:

$$t_{sleep(SP)} = \sup \left[ \frac{\sum_{i=0}^{i=n} T_{sleep}}{n} \right]^{0..n} \quad \forall n \in T \quad (5)$$

where n is the number of times the SP will be evaluated in the [0..T] interval, $T_{sleep}$ is the assigned sleep time according to eq. 4.

## IV. SIMULATION EXPERIMENTS AND DISCUSSION

In this section, we present the results extracted after conducting the simulation runs using the NS-2 [22] and the generated traffic traces for implementing the proposed scenario. The energy consumption model used in the simulation, for the calculation of the amount of energy consumed, is based theoretically on the WaveLAN PC/Card energy consumption characteristics found in study by Feeney and Nilsson [23]. These results are characterizing the trade-off issues of performance in deploying the discussed scenario for enabling EC through the SP's coordination and the proposed traffic-oriented scheme. Results also encompass comparisons with other existing schemes for the offered throughput, the reliability and the accuracy offered by the proposed scheme as well as EC efficiency conveying an estimated confidence interval (CI) of approximately 95% (mean values are shown in Figures 4-8). For every mean value extracted for each parameter at a simulation run of duration 18 hours each, a 95% confidence interval was evaluated. All confidence intervals were found to be less than 5% of the mean values of the certain extracted parameters. In this work the mobility model used is based on the probabilistic Fraction Brownian Motion (FBM) [24], where nodes are moving according to certain probabilities in accordance with the location and time. For implementing the described scenario, we used the spine model of [19] (based in C/Objective C programming language). Topology of a 'grid' based network was modeled according to the grid approach described in [14, 19]. Each node can directly communicate with other nodes if the area situated is in the same (3x3 center) rectangular area of the node. The network topology is generated according to [15, 19] where each node can communicate if it is set within a communicating "block". In the simulation of the proposed scenario we used a two-dimensional network, consisting of maximum 50 dense nodes located in each cluster. The topology changes dynamically as well as density and on a non-periodic basis (asynchronously as real time mobile users do). This issue enables us to examine different connectivity scenarios as well different self-tuning abilities of each node[1]. Each link (frequency channel) has max speed of 11Mb per sec (IEEE 802.11b in an Ethernet WLAN), and the propagation path loss is the two-ray model without fading. The traffic for mobile users is generated using the NS-2 [22] real-time traffic generator. Each device has different capacity considerations and battery utilization as in real-time the devices behave. This work also considers additional delay variation characteristics and delay-bounds in the end-to-end transmission path using signal strength and path-related characteristics as in [19]. The following results extracted, show the response of the proposed scheme using different evaluation angles, focusing in particular to the apparatus exhibited by the proposed traffic-aware mechanism and the EC efficiency provided by the scheme.

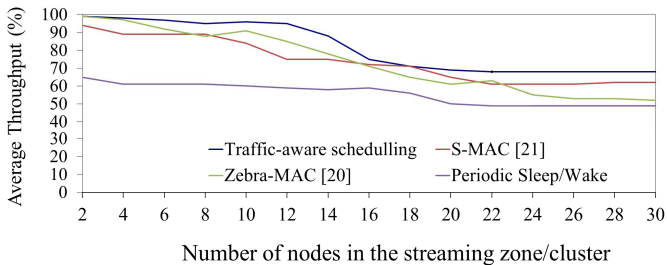

Figure 4. Average throughput in the streaming cluster.

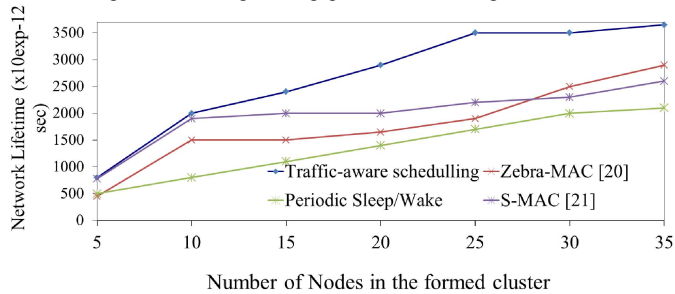

Figure 5. Comparison of existing schemes in contrast to the network intra-cluster lifetime.

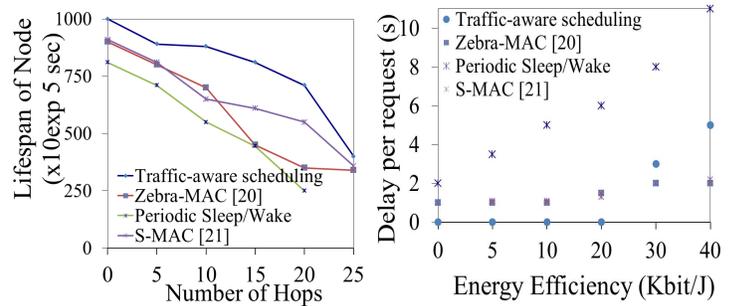

Figure 6. Lifespan of each node with the number of hops and the Energy efficiency with the delay.

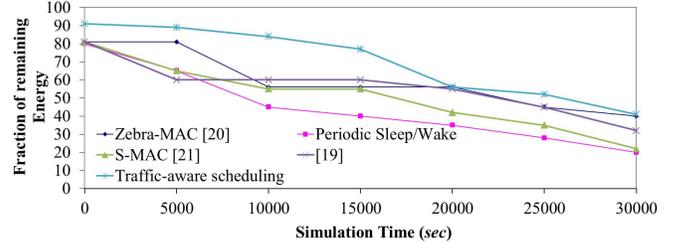

Figure 7. Fraction of the remaining Energy comparisons for different EC schemes.

The mechanisms hosted by the middleware try to maximize the availability of data and the tolerance for undeliverable packets by providing to users, access to replicas located in nearby neighboring hop-nodes. Moreover these mechanisms minimize the EC on each device and the EC consumed in the cluster as results show. Different sets of experiments were conducted comparing different priority schemes and their behavior under the same traffic and the same network dimensions.

The experimental phase aims to examine whether the proposed scheme could be able to cope in the presence of delay sensitive traffic (under bounded delay limitations) when traversing nodes with low residual energy and low remaining capacity. Figure 4 shows the comparative results of the proposed framework with other existing schemes such as Z-MAC, Periodic sleep/wake and S-MAC. Traffic-aware activity scheduling scheme shows to be more robust providing higher throughput in average compared to the existing schemes. Figure 5 shows the existing schemes in contrast to the network intra-cluster lifetime, whereas in Figure 6 a quantitative evaluation of the Lifespan of each node with the number of hops and the Energy efficiency with the delay id presented. Figure 6 shows the impact of each of the compared schemes in extending the lifetime of nodes. The proposed scheme showed supremacy against the Zebra-MAC and S-MAC schemes in terms of delay and lifespan extensibility. Figure 7 shows a fraction of the remaining energy in contrast to the responsiveness in EC for different schemes. Figure 8 presents the end-to-end path delay in accordance with the network's overall EC (μW) with the number of nodes for different energy-aware scheduling schemes. It is important to note that the EC is significantly improved in regards to the S-MAC and Zebra-MAC schemes where the consumption of each device has been reduced at a mean of 26% per cluster.

---

[1] Based on the remaining energy of each node in time.

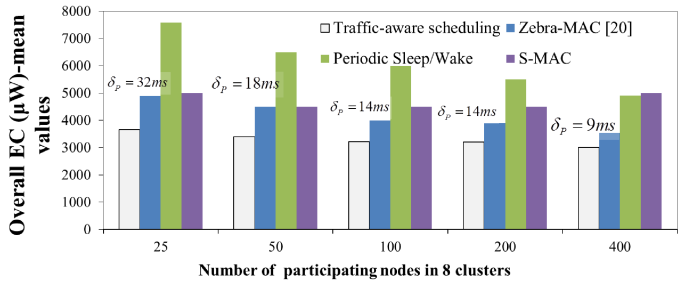

Figure 8. Network's overall EC (µW) with the number of nodes.

## V. CONCLUSIONS

This work proposes a reflective traffic-based mechanism taking into consideration the pairwise active-time comparison and backward traffic-difference characterization. Through the SPs sleep nodes are assigned and evaluated in a formed cluster taking into consideration the node's inter-cluster participation and the amount of time a node sleeps, in order to evaluate the next sleep time duration. This is performed through the proposed traffic-aware model, where the energy of each node can be tuned according to the activity periods and the traversed traffic, over a period of time, for a node. The performance evaluation through simulation shows that the proposed mechanism is manipulating the energy consumption of each device effectively, whereas the traffic-aware management scheme can explicitly reduce the energy consumed and, at the same time, keep the throughput response of the system at high levels. By comparing the extracted simulation results with other existing EC schemes, focusing on the performance evaluation and the EC offered, a significant supremacy of the proposed scheme is depicted with respect to the efficiency and the throughput response. The extracted results show that the proposed traffic-aware scheme uses optimally the network's and system's resources in terms of capacity and offers optimized EC and high throughput response.

ACKNOWLEDGMENT

We would like to thank the European FP7 COST Action IC0703 "Data Traffic Monitoring and Analysis (TMA): theory, techniques, tools and applications for the future networks", for the active support and cooperation.

REFERENCES

[1] Skype (2003) Skype Homepage. Available at http://www.skype.com/ [last accessed on March 2011].

[2] JXTA Chess (2003) JXTA Chess Homepage. Available at http://chess.jxta.org, [last accessed on March 2011].

[3] G. Kortuem, "Proem: a middleware platform for mobile peer-to-peer computing", Mobile Computing and Communications Review, ACM, Volume: 6, Issue: 4, October 2002, pp. 62-64.

[4] N. Kotilainen, M. Weber, M. Vapa, and J. Vuori, "Mobile Chedar - a Peer-to-Peer middleware for mobile devices," in Pervasive Computing and Communications Workshops, PerCom 2005 Workshops. Third IEEE International Conference on, 8-12 March 2005, pp. 86–90.

[5] C. Mascolo, L. Capra, S. Zachariadis, and W. Emmerich, "XMIDDLE: A Data-Sharing Middleware for Mobile Computing", International Journal on Wireless Personal Communications, Kluwer Academic Publisher, Volume: 21, Issue: 1, April 2002, pp. 77-103.

[6] M. Haahr, R. Cunningham, and . Cahill, (1999) Supporting CORBA Applications in a Mobile Environment (ALICE). 5th International Conference on Mobile Computing and Networking (MobiCom), August 1999; ACM Press, Seattle, WA.

[7] Joseph, A. D., Tauber, J. A., and Kaashoek, M. F. (1997) Mobile Computing with the Rover Toolkit. IEEE Transactions on Computers, 46(3).

[8] Schill, A., Bellmann, B., Bohmak, W., and Kummel, S. (1995) System support for mobile distributed applications. Proceedings of 2nd International Workshop on Services in Distributed and Networked Environments (SDNE), pp. 124-131; IEEE Computer Society Press, Whistler, British Columbia.

[9] M. Papadopouli, and H. Schulzrinne, "Design and Implementation of a Peer-to-Peer Data Dissemination and Prefetching Tool for Mobile Users", First New York Metro Area Networking Workshop, IBM T. J. Watson Research Center, Hawthorne, New York, 12 March 2002.

[10] R. Rajkumar, K. Juvva, A. Molano, and S. Oikawa, Resource Kernel: A Resource-Centric Approach to Real-Time Systems. Proceedings of the SPIE/ACM,Conference on Multimedia Computing and Networking, ACM, 1998.

[11] I. Jawhar, J. Wu, P. Agrawal, Resource Scheduling in Wireless Networks Using Directional Antennas, appears in: IEEE Transactions on Parallel and Distributed Systems, Sept. 2010, Volume: 21 Issue:9, pp. 1240 - 1253.

[12] S. Banerjee, A. Misra, "Minimum Energy Paths for Reliable Communication in Multi-hop Wireless Networks", ACM Symposium on Mobile Adhoc Networking and Computing (MOBIHOC 2002), Lausanne, Switzerland, June 9-11 2002, pp.146-156.

[13] S. Doshi and T. Brown. Minimum Energy Routing Schemes for a Wireless Ad Hoc Network, IEEE InfoCom 2002, New York, NY, June 23-27 2002, pp.113-119.

[14] C. X. Mavromoustakis, and H. D. Karatza, "Quality of Service Measures of Mobile Ad Hoc Wireless Network using Energy Consumption Mitigation with Asynchronous Inactivity Periods", Simulation: Transactions of the Society for Modelling and Simulation International, Volume 83, 2008, pp. 107-122.

[15] C. X. Mavromoustakis, "On the impact of caching and a model for storage-capacity measurements for energy conservation in asymmetrical wireless devices", IEEE Communication Society (COMSOC), 16th International Conference on Software, Telecommunications and Computer Networks (SoftCOM 2008), September 25 & 26 2008, "Dubrovnik", September 27, Split and Dubrovnik, pp. 243-247.

[16] W. R. Heinzelman, J. Kulik, and H. Balakrishnan. Adaptive protocols for information dissemination in wireless sensor networks. In Proceedings of MobiCom'99, pp. 174–185, Seattle, WA, USA, August 1999.

[17] O. Younis and S. Fahmy. Distributed clustering in ad-hoc sensor networks: a hybrid, energy-efficient approach. In Proceedings of IEEE INFOCOM 2004, Hong Kong, China, March 2004.

[18] C. X. Mavromoustakis, and H. D. Karatza, "A Gossip-based optimistic replication for efficient delay-sensitive streaming using an interactive middleware support system", IEEE Systems Journal, IEEE USA, Vol. 4, no. 2, p. 253-264, June 2010.

[19] C. X. Mavromoustakis, and H. D. Karatza, "Real time performance evaluation of asynchronous time division traffic-aware and delay-tolerant scheme in ad-hoc sensor networks", International Journal of Communication Systems (IJCS), Wiley, Volume 23 Issue 2 (February 2010), pp. 167-186.

[20] A. Rhee, M. Warrier, M. Aia, and J. Min. Z-MAC: a hybrid MAC for wireless sensor networks. In Proceedings of ACM SenSys'05, pp. 90–101, San Diego, CA, USA, 2005.

[21] W. Ye, J. Heidemann, and D. Estrin. Medium access control with coordinated adaptive sleeping for wireless sensor networks. IEEE/ACM Transactions on Networking, 12(3):493–506, June 2004.

[22] NS-2 Simulator, at http://www.isi.edu/nsnam/ns/.

[23] L. Feeney and M. Nilsson. Investigating the energy consumption of a wireless network interface in an ad hoc networking environment. In proc. of IEEE InfoCom, 5(8), 2001.

[24] C.X. Mavromoustakis and C.D. Dimitriou "Using Social Interactions for Opportunistic Resource Sharing using Mobility-enabled contact-oriented Replication", To appear in the proceeding of the 2012 International Conference on Collaboration Technologies and Systems (CTS 2012), Internet of Things, Machine to Machine and Smart Services Applications (IoT 2012), May 2012, Denver, USA.